\documentclass[a4paper,11pt]{article}
\usepackage{pos}
\usepackage{physics} 
\usepackage{CJKutf8}
\usepackage{subcaption}
\usepackage{graphicx}
\graphicspath{{./images/}}

\newcommand{\nn}{\nonumber\\}

\title{The lattice extraction of the TMD soft function using the auxiliary field representation of the Wilson line}
\ShortTitle{The lattice extraction of the TMD soft function}

\author[a]{Anthony Francis}
\author[b]{Issaku Kanamori}
\author[a,c,d]{C.-J. David Lin}
\author*[a]{Wayne Morris}
\author[e]{Yong Zhao}

\affiliation[a]{Institute of Physics, 
  National Yang Ming Chiao Tung University,\\
  1001 Ta-Hsueh Road, Hsinchu 30010, Taiwan}

\affiliation[b]{RIKEN Center for Computational Science,\\
  7-1-26 Minatojima-minami-machi, Chuo-ku, Kobe, Hyogo 650-0047, Japan}

\affiliation[c]{Centre for High Energy Physics, Chung-Yuan Christian University,\\
  Chung-Li 32023, Taiwan}

\affiliation[d]{Centre for Theoretical and Computational Physics, National Yang Ming Chiao Tung University,\\
  1001 Ta-Hsueh Road, Hsinchu 30010, Taiwan}

\affiliation[e]{Physics Division, Argonne National Laboratory,\\
  9700 S. Cass Avenue, Lemont, IL 60439, United States}

\emailAdd{waynemorris@nycu.edu.tw}
\abstract{
  The transverse momentum dependent (TMD) soft function can be obtained by formulating the Wilson line in terms of auxiliary 1-dimensional fermion fields on the lattice. 
  In this formulation, the directional vector of the auxiliary field in Euclidean space has the form $\tilde n = (in^0, \vec 0_\perp, n^3)$, where the time component is purely imaginary. 
  The components of these complex directional vectors in the Euclidean space can be mapped directly to the rapidities of the Minkowski space soft function. 
  We present the results of the one-loop calculation of the Euclidean space analog to the soft function using these complex directional vectors.
  As a result, we show that the calculation is valid only when the directional vectors obey the relation: $|r| = |n^3/n^0| > 1$, and that this result corresponds to a computation in Minkowski space with space-like directed Wilson lines.
  Finally, we show that a lattice calculable object can be constructed that has the desired properties of the soft function.
}

\FullConference{The 40th International Symposium on Lattice Field Theory (Lattice 2023)\\
July 31st - August 4th, 2023\\
Fermi National Accelerator Laboratory\\}

\begin{document}
\maketitle

%% Outline
%% 1. Introduction

\section{Introduction}

  Formulating the soft function on the lattice is a critical step in the investigation of TMD physics from lattice computations.
  The rapidity divergences present in computations of TMD objects, including the soft function, lead to unique challenges on the lattice, since the analog to rapidity in Euclidean space involves a Wick rotation. 
  %Motivated by a similar proposal in \cite{jiliuliu2020},
  In the case of the soft function, we propose to perform the lattice computation of the soft function with Wilson lines whose directional vectors take a purely imaginary time component, i.e. $\tilde n = (in^0,\vec 0_\perp, n^3)$.
  
  Using this formulation of the Wilson line would allow for a direct analytic continuation of the Euclidean soft function to its counterpart in Minkowski space in the space-like regime, since the soft function is time-independent. 
  We implement the Wilson line on the lattice through the auxiliary field representation, whose equation of motion can be solved iteratively in Euclidean time.
  There are further subtleties and complications related to the lattice realization of an auxiliary field propagator that will be discussed later in this paper.
  
  We will demonstrate that there is a direct connection between this complex Wilson line direction on the Euclidean lattice and the Minkowski space rapidity defined in Collins regularization scheme \cite{collins2011a}.
  In other words, our lattice computation corresponds to the Minkowski space soft function defined with space-like directed Wilson lines.
  %A commonly used definition of the soft function \cite{collins2011a} employs space-like directed Wilson lines in order to regulate rapidity divergences.
  
  %Collins' definition \cite{collins2011a} is the natural choice for the soft function obtained through the auxiliary field method, because it regulates rapidity divergences with Wilson lines pointing in space-like directions.

  When the Wilson line is pointing in a time-like direction, its auxiliary field representation corresponds to heavy quark effective theory (HQET). 
  In this case, the directional vector of the Wilson line corresponds to the heavy quark velocity. 
  It was proposed in \cite{jiliuliu2020} to calculate the soft function using HQET, 
  modeling the soft function as the form factor of a heavy quark pair. 
  In this formulation, the heavy quark velocity corresponds to the time-like directed Wilson lines in Minkowski space.
  Our work is motivated by this proposal, but differs in our use of Euclidean directional vectors that map to space-like directions in Minkowski space. Furthermore, we find that a Euclidean space computation of the soft function with directional vectors corresponding to time-like directions in Minkowski space gives a divergent integral.

  This paper will proceed as follows. 
  First, we will provide a brief review of TMD factorization in order to put our work into context. 
  Then, we will go into more detail on the theoretical motivation, including a perturbative analysis of the Euclidean space soft function computed using the complex Wilson lines. 
  After that, we will discuss the strategy for numerical implementation of this method.

\section{Review of TMD factorization}

  A comprehensive review of TMD factorization may be found in \cite{Boussarie:2023izj}, here we restrict ourselves to Drell-Yan scattering. 

  The Drell-Yan process involves the collision of two protons and produces a lepton pair in the final state: $pp \to \gamma^*/Z \to \ell^+ \ell^-$. 
  Here, measured final state variables are: $Q$, the invariant mass, $Y$, the rapidity, and $\vec q_\perp$, the transverse momentum.
  %In the region where $|\vec q_\perp| \sim \Lambda_{\rm QCD}$ {\color{red}YZ: $|\vec q_\perp| \sim \Lambda_{\rm QCD}$ is not necessary; the key hierarchy here is $|\vec q_\perp| \ll Q$, and of course $Q\gg \Lambda_{\rm QCD}$ for perturbation theory}, 
  The cross section may be factorized as:
  \begin{align}
    \frac{\dd \sigma}{\dd Q \dd Y \dd^2 \vec q_\perp}
    &=
    \sum_{i,j} H_{ij}\left(Q,\mu\right)
    \int \dd^2 \vec b_\perp e^{i\vec b_\perp \cdot \vec q_\perp}
    B_i\left(x_a, \vec b_\perp, \mu, \frac{\zeta_a}{\nu^2}\right)
    B_j\left(x_b, \vec b_\perp, \mu, \frac{\zeta_b}{\nu^2}\right)
    {S_{ij}\left(b_\perp,\mu,\nu\right)}
    \nn & \quad \times
    \left[ 
      1 + \mathcal O \left( \frac{q^2_\perp}{Q^2}, 
            \frac{\Lambda^2_{\rm QCD}}{Q^2} \right)
    \right] 
        \, ,
    \label{eq:DY1}
  \end{align}
  where the indices $i,j$ represent the quark flavor or gluon. The soft function is quark flavor independent, so one can write $S_{ij} \equiv S_i$.
  Here, $\mu$ is the renormalization scale, and $\zeta_{a,b}$ is the Collins-Soper (CS) scale. The rapidity scale, $\nu$ cancels between the beam and soft function, and $\zeta_a\zeta_b = Q^4$.
  The variables $x_{a,b}$ are the Bjorken-$x$ variables associated with the momentum fraction of the relevant parton inside of the proton.
  The transverse separation $\vec b_\perp$ is the coordinate space conjugate to $\vec q_\perp$. 
  Equation (\ref{eq:DY1}) is written in terms of the renormalized beam and soft functions, $B_i$ and $S_{i}$, which can be obtained from the bare functions through:
  \begin{align}
    B_i\left(x,\vec b_\perp, \mu, \zeta/\nu^2\right)
    &= 
    \lim_{\epsilon\to 0 }
      \lim_{\tau\to 0}
      Z^i_B\left(b_\perp, \mu, \nu, \epsilon, \tau, xP^+\right)
      \frac{B_i^{0(\rm u)}\left(x, \vec b_\perp, \epsilon, \tau, xP^+\right)}{S^{0(\rm subt)}_i\left(b_\perp, \epsilon, \tau\right)} 
      \label{eq:bren} \\
    S_i\left(b_\perp, \mu, \nu \right)
    &= 
    \lim_{\epsilon\to 0 }
      \lim_{\tau\to 0}
      Z^i_S\left(b_\perp, \mu, \nu, \epsilon, \tau\right)
      S^0_i\left(b_\perp, \epsilon, \tau\right) ,
      \label{eq:sren}
  \end{align}
  where $Z^i_B$ and $Z^i_S$ are the UV renormalization factors, and $\epsilon$ is a parameter associated with dimensional regularization.
  The term in the denominator of Eq. (\ref{eq:bren}), $S^{0(\rm subt)}_i$, is introduced to account for the overlap of infrared (IR) regions between the beam and soft function. As such, it takes care of possible double counting. 
  
  The beam functions capture long-distance physics associated with the incoming protons, while the soft function accounts for soft gluon radiation in the final state. It is not possible to directly measure the soft function experimentally, and so it must be calculated via non-perturbative methods.

  A central aspect of TMD factorization are the rapidity divergences associated with partons moving in the opposite direction of their parent hadron with infinite rapidity.
  Because rapidity divergences are not associated with IR physics nor ultraviolet (UV) divergences, special care must be taken in regulating them.
  Above, $\tau$ represents a generic rapidity regulator, and $\nu$ the associated scale. There are a number of possible regulators to choose from, including the one introduced by Collins \cite{collins2011a}. Being an unphysical divergence, the dependence on the rapidity scale necessarily cancels between the two beam functions and the soft function.

  Choosing the Collins scheme in place of the generic rapidity regulator, $\tau$, the beam and soft functions are defined as:
  \begin{align}
    B^{0(u)}_i\left(x,\vec b_\perp, \epsilon, y_B, xP^+\right)
    & =
      \int \frac{\dd b^-}{2\pi} e^{-ib^-(xP^+)}
      \bra{P} 
        \bar \psi^0_i\left(b^-, \vec b_\perp\right)
        W_{n_B}\left(b^-,\vec b_\perp;-\infty,0\right)
    \nn & \qquad \quad \times \
        W_\perp\left(-\infty n_B; 0, b_\perp \right)
        W_{n_B} \left(0; 0, -\infty\right)
        \frac{\gamma^+}{2}
        \psi^0_i\left(0\right)
      \ket{P} \, , 
    \\
    S^0\left(b_\perp, \epsilon, y_A - y_B\right)
    &= \frac 1{N_c} 
      \bra{0} 
        W_{n_A}\left(b_\perp;0,-\infty\right) 
        W_{n_B}\left(b_\perp;-\infty,0\right)
        W_\perp\left(-\infty n_B;0,b_\perp\right)
    \nn & \qquad \qquad 
        W_{n_B}\left(0;0,-\infty\right) 
        W_{n_A}\left(0;-\infty,0\right)
        W_\perp\left(-\infty n_A;b_\perp,0\right)
       \ket{0} \, ,
    \label{eq:softCollins}
  \end{align}
  where
  \begin{align}
    W_n\left(x;a,b\right)
    &= P \exp\left\{
      -ig_0 \int_a^b \dd s \, 
      n^\mu A_\mu^{c0}\left(x+sn\right) t^c 
    \right\} \, .
  \end{align}
  The beam and soft functions are naively defined with Wilson lines pointing in light-like directions, defined by directional vectors $n=(1,0,0,1)$, and $\bar n = (1,0,0,-1)$. In the Collins scheme, however, the Wilson lines now point in space-like directions, and we make the replacements:
  \begin{align}
    n \to n_A = n - e^{-2y_A}\bar n, 
    \qquad \bar n \to n_B = \bar n - e^{2y_B} n \, ,
    \label{spaceDir}
  \end{align}
  where $y_A$ and $y_B$ are the rapidities of the Wilson lines, and the light-cone direction is recovered in the limits $y_A\to\infty$ and $y_B\to-\infty$.
  The TMDPDF can then be constructed from a combination of the beam and soft function:
  \begin{align}
    f_i (x, \vec b_\perp, \mu, \zeta )
    & = 
      \lim_{\epsilon\to 0}
      Z^i_{\rm UV}\left(\mu, \epsilon, \zeta\right)
    \nn & \quad \times
      \lim\limits_{\substack{y_A\to+\infty \\ y_B\to -\infty}}
      B_i\left(x,\vec b_\perp, \epsilon, y_B, xP^+\right)
      \sqrt{
        \frac{S_i\left(b_\perp,\epsilon,y_A-y_n\right)}
        {
          S_i\left(b_\perp,\epsilon,y_A-y_B\right)
          S_i\left(b_\perp,\epsilon,y_n-y_B\right)
        }
      } \, .
  \end{align}
  where $\zeta = 2(xP^+)^2e^{-2y_n}$ is the CS scale \cite{collins2011a} for the $n$-collinear proton. 
  %Taking the product of $\zeta_a$, with the CS scale for the $\bar n$-collinear proton, $\eta_b = 2(x_b P^-_B)^2e^{2y_n}$, we get the invariant mass of the hard scattering process.

  The TMDPDF, along with the beam and soft functions, follows a set of evolution equations in the renormalization and CS scale. 
  The equation relevant to our discussion is:
  \begin{align}
    %\dv{\ln f_{i} \left(x, \vec b_\perp, \mu, \zeta \right)}
    %  {\ln \mu}
    %=
    %\gamma^q_\mu\left(\mu, \zeta\right) \, , \quad
    \pdv{\ln f_{q} \left(x, \vec b_\perp, \mu, \zeta \right)}
      {\ln \sqrt{\zeta}}
    =
      \gamma^q_\zeta \left(\mu, b_\perp\right) \, . %\quad 
      \label{eq:csk}
    %\dv{\gamma^q_\zeta \left(\mu, b_\perp\right)}{\ln \mu}
    %=
    %-2\Gamma^q_{\rm cusp} \left[ \alpha_s(\mu) \right] \, ,
  \end{align}
  where we have chosen $i=q$ for the quark TMDPDF.
  The CS kernel, $\gamma^q_\zeta \left(\mu, b_\perp\right)$, governs the evolution of the TMDPDF with respect to the change of the rapidity scale, $\zeta$.
  %Eq. (\ref{eq:csk}) gives the CS kernel, which governs the rapidity evolution of the TMDPDF. 
  It can be obtained direactly from the beam or the soft function.  For instance,
  %The CS kernel may also be obtained directly from the beam or soft function:
  \begin{align}
    %\zeta_a = \left( x m_P e^{y_P-y_n} \right)^2, \qquad
    \gamma_\zeta^q (\mu, b_\perp) 
    %= \dv{\ln B_q}{y_P} 
    = \dv{\ln S_q\left(b_\perp,\epsilon,y\right)}{y}
    - {\rm UV~counterterms} \, ,
  \end{align}
  and so, by putting this formalism on the lattice, it is possible to determine the Collins-Soper kernel as well as the soft function from one lattice calculation.
  %in addition to the soft function, it is possible to determine the Collins-Soper kernel from one lattice calculation.

  The soft function and the CS kernel are also necessary components in the calculation of TMDPDFs on the lattice. 
  Currently, there is a proposal for the extraction of TMDPDFs using the quasi-PDF approach (\cite{Ji:2014hxa,Ji:2018hvs,Ebert:2019okf,jiliuliu2020,Ebert:2022fmh}). 
  The matching relation connecting the quasi-TMDPDF, $\tilde f_q$, to the light-cone TMDPDF in the case of unpolarized quarks is:
  \begin{align}
    \tilde f_{q}\left(x, \vec b_\perp, \mu \tilde, \zeta, x\tilde P^z \right)
    &=
    C_q\left(x\tilde P^z, \mu\right) 
    \exp\left[\frac 12 \gamma^q_\zeta\left(\mu, b_\perp\right) \log \frac{\tilde \zeta}{\zeta}\right]
    f_{q}\left(x, \vec b_\perp, \mu, \zeta\right) \nn
    & \quad \times
    \left\{
      1 + 
      \mathcal{O}
        \left(
          \frac 1{(x\tilde P^z b_\perp)^2}, 
          \frac{\Lambda_{\rm QCD}^2}{(x\tilde P^z)^2}
        \right)
    \right\} \, ,
    \label{eq:match}
  \end{align}
  where $C_q$ is a perturbatively calculable matching kernel, and $\tilde \zeta = x^2m_h^2 e^{2(y_{\tilde P}+y_B-y_n)}$.
  The quasi-TMDPDF is constructed as~\cite{Ebert:2022fmh}:
  \begin{align}
    \tilde f_q \left( x, \vec b_\perp, \mu, \tilde \zeta, x\tilde P^z \right)
    &=
    \tilde f_q^{\rm naive} \left( x, \vec b_\perp, \mu \tilde \zeta, x\tilde P^z \right)
    \sqrt{\frac{\tilde S_q^{\rm naive}\left(b_\perp, \mu\right)}{S_q\left(b_\perp,\mu,2y_n,2y_B\right)}} \, ,
    \label{eq:naive}
  \end{align}
  where $\tilde f_q^{\rm naive}$ and $\tilde S_q^{\rm naive}$ are lattice calculable objects, and $S_q$ is the Collins soft function. 
  The lattice extracted Collins soft function and CS kernel can be used in obtaining $f_q$ from 
  %We can substitute our result for the Collins soft function and CS kernel into 
  Eqs. (\ref{eq:match}) and (\ref{eq:naive}).

  An extraction of the TMDPDF was performed by the Lattice Parton Collaboration (LPC) \cite{he2022} using a different method \cite{jiliuliu2020} to obtain the soft function.

\section{Theoretical considerations}

  We perform our perturbative analysis in the case of infinite and finite length Wilson lines. 
  For infinite Wilson lines we demonstrate an exact correspondence between the Euclidean space and Minkowski space computation at one loop.
  Finally, we show that the finite length Wilson line computation approaches the infinite length result after constructing a ratio to cancel out the finite length effects.

\subsection{Infinite Wilson lines}
  \begin{figure}
    \centering
    \begin{subfigure}{0.24\textwidth}
      \includegraphics[width=\textwidth]{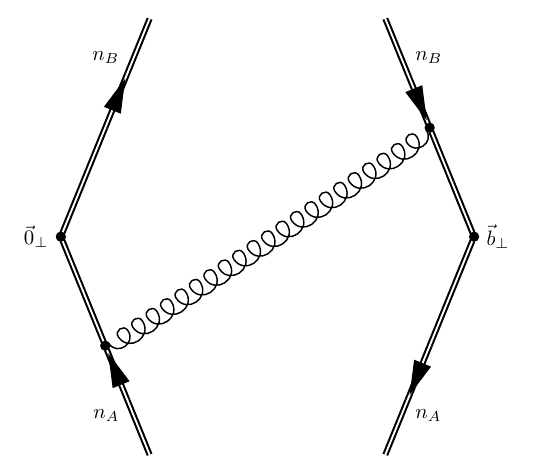}
      \caption{}
      \label{fig:Sa}
    \end{subfigure}
    \begin{subfigure}{0.24\textwidth}
      \includegraphics[width=\textwidth]{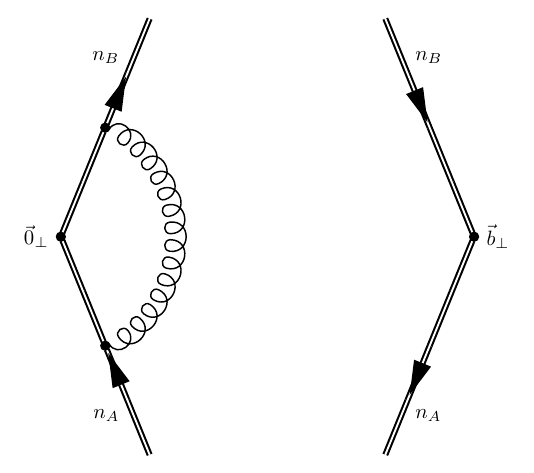}
      \caption{}
      \label{fig:Sb}
    \end{subfigure}
    \begin{subfigure}{0.24\textwidth}
      \includegraphics[width=\textwidth]{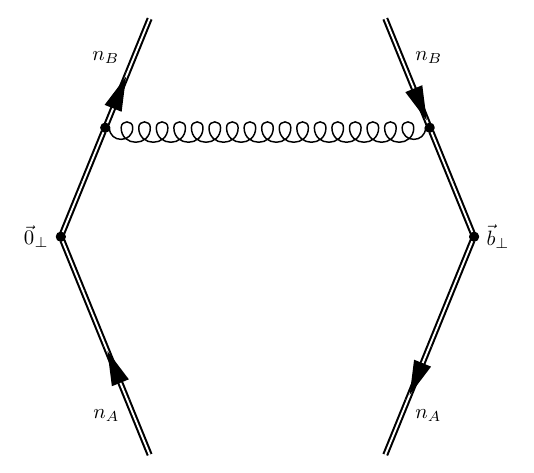}
      \caption{}
      \label{fig:Sc}
    \end{subfigure}
    \begin{subfigure}{0.24\textwidth}
      \includegraphics[width=\textwidth]{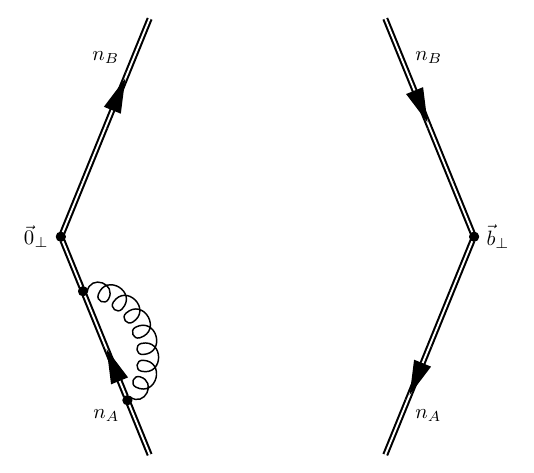}
      \caption{}
      \label{fig:Sd}
    \end{subfigure}
    \caption{Diagrams contributing at one-loop to the soft function}
    \label{fig:S}
  \end{figure}

  We first perform a one-loop computation of the soft function in Euclidean space with complex directions in order to make a connection with the Minkowski space result:
  \begin{align}
  \tilde n_A = (in_A^0, \vec 0_\perp, n_A^3), \qquad
  \tilde n_B = (in_B^0, \vec 0_\perp, -n_B^3) 
  \end{align}
  For convenience we define the ratios: $r_a = n_A^3/n_A^0$, and $r_b = n_B^3/n_B^0$.
  We perform the computation in d space-time dimensions in order to regulate UV divergences. Before presenting the full result, we can gain important insights by first looking at diagram a in figure \ref{fig:S}:
  \begin{align}
    S_{\ref{fig:Sa}}\left(b_T, \epsilon, r_a, r_b\right)
    %%%%%%%%%%%%%%%%%%%%%%%%% LINE 1 %%%%%%%%%%%%%%%%%%%%%%%%%
    &= i g^2 C_F  (\tilde n_{A,\mu} \tilde n_{B,\nu}) \int_{-\infty}^0 \dd s \int_{-\infty}^0 \dd t 
        A^\mu (b+s\tilde n_A) A^\nu ( t\tilde n_B) \nn
    %%%%%%%%%%%%%%%%%%%%%%%%% LINE 2 %%%%%%%%%%%%%%%%%%%%%%%%%
    &= i g^2 C_F  (\tilde n_A \cdot \tilde n_B) \int_{-\infty}^0 \dd s \int_{-\infty}^0 \dd t 
    \int \frac{\dd^d k}{(2\pi)^d} e^{-ik(b+s\tilde n_A-t\tilde n_B)} \frac{-i}{k^2} \, ,
    \label{eq:s1aa}
  \end{align}
  where on line 2 of Eq. (\ref{eq:s1aa}), we wrote the gluon propagator in Euclidean space. 
  From here, one usually proceeds with the computation in momentum space by first integrating over $s$ and $t$ to obtain the Wilson line propagator. However, these integrals are not well defined, since there is a finite real part in the exponential. Looking at the $s$ integration:
  \begin{align}
    \int_{-\infty}^0 \dd s \, e^{sn_A^0 k_4 - isn_A^3k_3}
    &=
    \left. 
      \frac{e^{sn_A^0 k_4 - isn_A^3 k_3}}
        {n_A^0 k_4 - in_A^3 k_3} 
    \right|^0_{s=-\infty}
    \overset{k_4<0}{\longrightarrow} \quad \infty \, .
  \end{align}
  Without a clear way to handle this integration, we instead approach the computation by performing the momentum integral in Eq. (\ref{eq:s1aa}) first. 
  In order to do this, we apply the Schwinger parametrization, and complete the square in $k$:
  \begin{align}
    S_{\ref{fig:Sa}}\left(b_T, \epsilon, r_a, r_b\right)
    %%%%%%%%%%%%%%%%%%%%%%%%% LINE 4 %%%%%%%%%%%%%%%%%%%%%%%%%
    &= g^2 C_F  (\tilde n_A \cdot \tilde n_B) \int_{-\infty}^0 \dd s \int_{-\infty}^0 \dd t \int_0^\infty \dd u
    \int \frac{\dd^d k}{(2\pi)^d} 
    e^{-uk^2} 
    e^{-(b+s\tilde n_A-t\tilde n_B)^2/4u} \nn
    %%%%%%%%%%%%%%%%%%%%%%%%% LINE 7 %%%%%%%%%%%%%%%%%%%%%%%%%
    %&= \frac{g^2 C_F}{(4\pi)^{d/2}}  (\tilde n_A \cdot \tilde n_B) \int_{-\infty}^0 \dd s \int_{-\infty}^0 \dd t 
    %\frac{\Gamma(d/2-1)}{\left( (b+s\tilde n_A-t\tilde n_B)^2/4\right)^{d/2-1}}
    %\nn
    %%%%%%%%%%%%%%%%%%%%%%%%%% LINE 10 %%%%%%%%%%%%%%%%%%%%%%%%%
    %&= \frac{g^2 C_F}{4\pi^{2-\epsilon}}  (\tilde n_A \cdot \tilde n_B) 
    %\int_{-\infty}^0 \lambda^{-1+2\epsilon} \dd \lambda \int_{1}^0 \dd x 
    %\frac{\Gamma(d/2-1)}{\left( - x^2 (a_4^2-a_3^2) 
    %                + 2 x \bar x(a_4b_4+a_3b_3) 
    %                - \bar x^2(b_4^2-b_3^2) + b_T^2/\lambda^2 \right)^{1-\epsilon}}
    %\nn
    %%%%%%%%%%%%%%%%%%%%%%%%%% LINE 11 %%%%%%%%%%%%%%%%%%%%%%%%%
    %&= \frac{g^2 C_F}{4\pi^{2-\epsilon}}   
    %\frac{ \left(b_T^2\right)^{\epsilon }\Gamma(1-\epsilon)}{2 \epsilon }
    %\int_{1}^0 \dd x 
    %\frac{(-a_4b_4 - a_3b_3)}
    %                {-x^2 (a_4^2-a_3^2) 
    %                + 2 x \bar x(a_4b_4+a_3b_3) 
    %                - \bar x^2(b_4^2-b_3^2)}
    %\nn
    %%%%%%%%%%%%%%%%%%%%%%%%% LINE 11.4 %%%%%%%%%%%%%%%%%%%%%%%%%
    &= \frac{g^2 C_F}{4\pi^{2-\epsilon}}   
    \frac{ \left(b_T^2\right)^{\epsilon }\Gamma(1-\epsilon)}{2 \epsilon_{\rm IR} }
    \frac 12 
    \log \left(\frac{\left(r_a-1\right) \left(r_b-1\right)}{\left(r_a+1\right) \left(r_b+1\right)} \right)
    \frac{r_a r_b+1}{ \left(r_a+r_b\right)} \, .
    \label{eq:s1ab}
  \end{align}
  Thus, we obtain a finite result for diagram \ref{fig:Sa}. However, this result is only valid when $|r_a|,|r_b|>1$. This can be seen from expanding the term in the exponential of line 1 in Eq. (\ref{eq:s1ab}):
  \begin{align}
    (b+s\tilde n_A-t\tilde n_B)^2
    &= b_\perp^2 + s^2 (n_A^0)^2\left( r_a^2 -1 \right)
      + t^2 (n_B^0)^2\left( r_b^2 -1 \right)
      + 2st n_A^0 n_B^0 \left( r_a r_b +1 \right) 
      > 0 \, .
      \label{rabIneq}
  \end{align}
  Equation (\ref{rabIneq}) must be positive in order for the integral in Eq. (\ref{eq:s1ab}) to converge, which is only true when the
  %The second and third terms on the RHS of Eq. (\ref{rabIneq}) ensure that the 
  absolute values of $r_a$ and $r_b$ are greater than one.
  %The consequences of the last term are discussed below.

  The full one-loop result in Euclidean space with complex directional vectors is:
  \begin{align}
    S(b_T,\epsilon,r_a,r_b)
    %%%%%%%%%%%%%%%%%%%%%%%%% LINE 1 %%%%%%%%%%%%%%%%%%%%%%%%
    &= 1 + \frac{\alpha_s C_F}{2\pi}    
    \left(\frac{1}{\epsilon } + \log (\pi b_\perp^2 \mu_0^2 e^{\gamma_E})\right)
    \left\{
      2
      +
      \log
      \left(
        \frac{\left(r_a-1\right)\left(r_b-1\right)}
          {\left(r_a+1\right)\left(r_b+1\right)}
      \right)
      \frac{r_a r_b+1}{r_a + r_b}
    \right\}
    \label{eq:ole}
  \end{align}

  Collins scheme uses space-like Wilson lines defined in Eq. (\ref{spaceDir}), but in principle one can also use time-like Wilson lines to regulate the rapidity divergence. For completeness, we list both space-like and time-like cases here:
  \begin{align}
    \text{Time-like:}& \quad 
      n_A = \left( 1+e^{-2y_A}, \vec 0_\perp, 1-e^{-2y_A}  \right)
      , \quad 
      n_B = \left( 1+e^{2y_B}, \vec 0_\perp, -1+e^{2y_B}  \right) \\
    \text{Space-like:}& \quad 
      n_A = \left( 1-e^{-2y_A}, \vec 0_\perp, 1+e^{-2y_A}  \right)
      , \quad 
      n_B = \left( 1-e^{2y_B}, \vec 0_\perp, -1-e^{2y_B}  \right) 
  \end{align}
  Because the complex directional vectors in Euclidean space are written in terms of the elements of their Minkowski space counterparts, we write $r_a$ and $r_b$ in terms of the components of the space-like or time-like directional vectors. For the time-like case, we get:
  \begin{align}
    r_a = \frac{1-e^{-2y_A}}{1+e^{-2y_A}}
    , \quad 
    r_b = \frac{1-e^{2y_B}}{1+e^{2y_B}} \, ,
  \end{align}
  where we can see that $-1<r_a,r_b<1$. We immediately find that this fails to satisfy the condition set by Eq. (\ref{rabIneq}). As for the space-like case, we find:
  \begin{align}
    r_a = \frac{1+e^{-2y_A}}{1-e^{-2y_A}}
    , \quad 
    r_b = \frac{1+e^{2y_B}}{1-e^{2y_B}} \, ,
    \label{eq:spaceR}
  \end{align}
  which satisfies Eq. (\ref{rabIneq}). 
  The third condition set by Eq. (\ref{rabIneq}) demands that $n_A^0 n_B^0 (r_a r_b +1) >0$, which indicates that the Wilson lines must be both future pointing or both past pointing, corresponding to $e^+e^-$ annihilation or DY type processes. 
  %Processes of SIDIS type apparently cannot be calculated with this method. {\color{red}YZ: The unpolarized TMD is process independent, so it is a necessity to calculate the soft function for SIDIS. Better not make a such a statement.}

  Substituting Eq. (\ref{eq:spaceR}) into Eq. (\ref{eq:ole}), we recover the Minkowski space result:
  \begin{align}
    S(b_\perp, \epsilon, y_A, y_B)
    &=
    1 + \frac{\alpha_s C_F}{2\pi}
    \left( \frac 1{\epsilon} 
      + \ln \left(\pi b_\perp^2 \mu_0^2e^{\gamma_E} \right) \right)
    \left\{
      2 -
      2 |y_A - y_B|
      \frac{1+e^{2(y_B-y_A)}}{1-e^{2(y_B-y_A)}}
    \right\}
    + \mathcal O(\alpha_s^2)
    %&=
    %1 + \frac{\alpha_s C_F}{2\pi}
    %\left( \frac 1{\epsilon} 
    %  + \ln \left(\pi b_\perp^2 \mu_0^2e^{\gamma_E} \right) \right)
    %(2 - 2 |y_A - y_B|) + \mathcal O(\alpha_s^2)
  \end{align}
  \subsection{Finite Wilson lines}

  The calculation on the lattice is limited by its finite space-time volume, and Wilson lines therefore have a finite length $L$.  To take this into account we also perform a one-loop calculation of finite length Wilson lines.
  One of the consequences of finite length Wilson lines is the presence of a linear divergence in $L$. We can remove the linear divergence by constructing the ratio (\cite{collins2011b,jiliuliu2020}):
  \begin{align}
    S\left(b_\perp,a,r_a,r_b\right)
    &=
    \lim_{L\to\infty}
    \frac{S\left(b_\perp,a,r_a,r_b,L\right)}
      {\sqrt{S\left(b_\perp,a,r_a,-r_a,L\right)S\left(b_\perp,a,-r_b,r_b,L\right)}} \, ,
    \label{eq:ratio}
  \end{align}
  where $a$ is some generic regulator used for the UV divergence. 
  %The two functions in the denominator of Eq. (\ref{eq:ratio}) amount to a rectangular Wilson loop of length $2L$.
  We have found that Eq. (\ref{eq:ratio}) holds to one loop in perturbation theory, using a Polyakov regulator for the UV.
  Additionally, we have found that power corrections of order $b_\perp^2/L^2$ should also cancel in the ratio.
  Computing the ratio in Eq. \ref{eq:ratio} on the lattice at large enough $L$ should produce a time-independent observable up to power suppressed corrections in $b_\perp^4/L^4$, since $L$ on the lattice will be proportional to Euclidean time.

\section{Strategy of numerical implementation}

  It is well known (\cite{gervaisnevau1979,arefeva1980}) that the Wilson line can be written in terms of a one-dimensional auxiliary fermion field that `lives' along the path of the Wilson line:
  \begin{align}
    P \exp 
      \left\{-ig\int_{s_i}^{s_f} \dd s n^\mu A_\mu(y(s)) \right\}
    &= 
    Z_{\psi}^{-1} \int \mathcal D \psi \mathcal D \bar \psi \,
      \psi \bar\psi
      \exp
      \left\{
        i \int_{s_i}^{s_f} \dd s \bar\psi i n\cdot\partial \psi 
        - g_0 \bar\psi n \cdot A \psi
      \right\} \, , 
  \end{align}
  whose propagator solves the Green function:
  \begin{align}
    in\cdot D H_n(x-y) = i\delta^{(4)}(x-y) \, ,
    \label{eq:grM}
  \end{align}
  with $D=\partial+ig_0 A$ the covariant derivative. 
  %When $n$ is time-like, this procedure corresponds to HQET.
  The Euclidean space analog of Eq. (\ref{eq:grM}) has a directional vector with a purely imaginary time component, and can be written as $\tilde n = (in^0,\vec n)$ in terms of the components of the Minkowski space vector. 
  After performing a Wick rotation, the Euclidean space Green function is:
  \begin{align}
    i \tilde n \cdot D_E H_{\tilde n} (x_E-y_E) = \delta^{(4)}(x_E-y_E) \, .
    \label{eq:greenE}
  \end{align}
  As argued in \cite{agliettieetal1992, Aglietti:1993hf}, meaningful solutions for Eq. (\ref{eq:greenE}) can only be obtained with a UV cutoff.
  %While Eq. (\ref{eq:greenE}) has no solution, we can define a modified Euclidean space propagator using some form of UV regulator.
  Therefore, with the lattice spacing serving as our regulator, a discretized version of this propagator can be constructed.
  The central idea then is to construct the soft function in terms of the lattice regulated versions of these auxiliary field propagators, $H_{\tilde n}$, which can then be shown to have a well-defined continuum limit.
  %whose Euclidean space equation of motion can be discretized and solved on the lattice
  
  Solving the auxiliary field propagator on the lattice proceeds in the exact same way as in HQET (\cite{Mandula:1990fit, mandulaogilvie1992, agliettieetal1992, Aglietti:1993hf}), but now we are no longer bound by the kinematical restrictions on heavy quarks. Therefore, we use the recursive relation given in Eq. (31) of \cite{mandulaogilvie1992} to construct the lattice auxiliary field propagators. Based on our perturbative analysis, we expect that computing Eq. (\ref{eq:ratio}) on the lattice will approach a time independent result at large enough Euclidean time.

\section{Conclusions}

  Extracting the soft function through the auxiliary field method shows promise based on our perturbative analysis, 
  given the time independence of the soft function, and the direct relationship between the complex directional vectors, $\tilde n_A, \tilde n_B$, in Euclidean space, and $y_A, y_B$ in Minkowski space. 
  An important point that we did not discuss here is that a direct analytic continuation of our lattice computation to Minkowski space may be complicated by the pole structure of the auxiliary field propagator. 
  However, an analysis of this problem was already done in the context of moving-HQET in \cite{Aglietti:1993hf}.
  %, which implies that problem may be solved through the construction of a perturbative matching relation. 
  Finally, the construction of the ratio in Eq. (\ref{eq:ratio}) solves the issues related to the linear divergence in $L$ and removes power corrections up to, at least, $b_\perp^2/L^2$. 
  We have started the numerical implementation of this formalism on the lattice. In the near future we will present first results with the aim of establishing and demonstrating the feasibility of the presented ideas.
  \footnote{ 
  After this proceeding was published, we were made aware of a lecture note~\cite{Liu:2022nnk} which implies that one can analytically continue the heavy-quark form factor~\cite{jiliuliu2020} in the $z$-axis to obtain Euclidean Wilson lines along the $(n^0,0,0,in^3)$ direction. This would be equivalent to our proposal through O(4) symmetry~\cite{Liu:2022nnk} if $n^3<n^0$. Nevertheless, such an analytical continuation was not explicitly pointed out in Refs.~\cite{jiliuliu2020,Liu:2022nnk}, nor was there mention of the failure of the standard timelike moving HQET in calculating the soft function, which was discovered in this work.
  %After this proceeding was published, we were made aware of a lecture note~\cite{Liu:2022nnk} which implies that one can analytically continue the heavy-quark form factor~\cite{jiliuliu2020} in the $z$-axis to obtain Euclidean Wilson lines along the $(n^0,0,0,in^3)$ direction. This would be equivalent to our proposal through O(4) symmetry~\cite{Liu:2022nnk} if $n^3<n^0$. Nevertheless, such an analytical continuation was not pointed out in Ref.~\cite{jiliuliu2020}, nor was there mention of the failure of the standard timelike moving HQET in Euclidean space, which was discovered in this work.
  %After publishing this paper we found a lecture note \cite{liu2022} that discusses the analyticity of the soft function as a function of the hyperbolic angle between Wilson line directions.
  } 

\acknowledgments

  We acknowledge Yizhuang Liu for helpful discussions.
  AF is supported by the National Science and Technology Council (NSTC) of Taiwan under grant 111-2112-M-A49-018-MY2. CJDL is supported by 112-2112-M-A49-021-MY3, and WM is supported by 112-2811-M-A49-517-MY2 and 112-2112-M-A49-021-MY3 from NSTC. YZ is supported by the U.S. Department of Energy, Office of Science, Office of Nuclear Physics through Contract No.~DE-AC02-06CH11357, and the 2023 Physical Sciences and Engineering (PSE) Early Investigator Named Award program at Argonne National Laboratory.


\begin{thebibliography}{99}

\bibitem{collins2011a}
J. Collins,
\emph{Foundations of perturbative QCD}, 
Cambridge monographs on particle physics, nuclear physics, and cosmology. Cambridge Univ. Press, New York, NY, 2011
.

\bibitem{jiliuliu2020}
X. Ji, Y.Z. Liu and Y.S. Liu,
\emph{TMD soft function from large-momentum effective theory},
\href{https://doi.org/10.1016/j.nuclphysb.2020.115054}
  {\emph{Nucl. Phys. B} \textbf{955} (2020) 115054},
[{\tt hep-ph/1910.11415}]
.

%\cite{Boussarie:2023izj}
\bibitem{Boussarie:2023izj}
R.~Boussarie, M.~Burkardt, M.~Constantinou, W.~Detmold, M.~Ebert, M.~Engelhardt, S.~Fleming, L.~Gamberg, X.~Ji and Z.~B.~Kang, \textit{et al.}
\emph{TMD Handbook},
\href{https://doi.org/10.48550/arXiv.2304.03302}
{[{\tt arXiv:2304.03302 [hep-ph]}]}
.
%31 citations counted in INSPIRE as of 17 Nov 2023

%\cite{Ji:2014hxa}
\bibitem{Ji:2014hxa}
X.~Ji, P.~Sun, X.~Xiong and F.~Yuan,
\emph{Soft factor subtraction and transverse momentum dependent parton distributions on the lattice},
\href{https://doi.org/10.1103/PhysRevD.91.074009}
{\emph{Phys. Rev. D} \textbf{91}, 074009 (2015)},
%doi:10.1103/PhysRevD.91.074009
[{\tt arXiv:1405.7640 [hep-ph]}]
.
%102 citations counted in INSPIRE as of 03 Dec 2023

%\cite{Ji:2018hvs}
\bibitem{Ji:2018hvs}
X.~Ji, L.~C.~Jin, F.~Yuan, J.~H.~Zhang and Y.~Zhao,
\emph{Transverse momentum dependent parton quasidistributions},
\href{https://doi.org/10.1103/PhysRevD.99.114006}
{\emph{Phys. Rev. D} \textbf{99}, no.11, 114006 (2019)},
%doi:10.1103/PhysRevD.99.114006
[{\tt arXiv:1801.05930 [hep-ph]}].
%65 citations counted in INSPIRE as of 03 Dec 2023

%\cite{Ebert:2019okf}
\bibitem{Ebert:2019okf}
M.~A.~Ebert, I.~W.~Stewart and Y.~Zhao,
\emph{Towards Quasi-Transverse Momentum Dependent PDFs Computable on the Lattice},
\href{https://doi.org/10.1007/JHEP09(2019)037}
{\emph{JHEP} \textbf{09}, 037 (2019)},
%doi:10.1007/JHEP09(2019)037
[{\tt arXiv:1901.03685 [hep-ph]}]
.
%91 citations counted in INSPIRE as of 29 Nov 2023

%\cite{Ebert:2022fmh}
\bibitem{Ebert:2022fmh}
M.~A.~Ebert, S.~T.~Schindler, I.~W.~Stewart and Y.~Zhao,
\emph{Factorization connecting continuum \& lattice TMDs},
\href{https://doi.org/10.1007/JHEP04(2022)178}
{\emph{JHEP} \textbf{04}, 178 (2022)},
%doi:10.1007/JHEP04(2022)178
[{\tt arXiv:2201.08401 [hep-ph]}]
.
%28 citations counted in INSPIRE as of 29 Nov 2023


\bibitem{he2022}
J. C. He, \textit{et al.}
\emph{Unpolarized Transverse-Momentum-Dependent Parton Distributions of the Nucleon from Lattice QCD},
\href{https://doi.org/10.48550/arXiv.2211.02340}
{[{ \tt arXiv:2211.02340 [hep-lat]}]}
.

\bibitem{collins2011b}
J. Collins,
\emph{New definition of TMD parton densities},
\href{https://doi.org/10.1142/S2010194511001590}
  {\emph{Int. J. Mod. Phys. Conf. Ser. 4} (2011) 85-96},
[{\tt hep-ph/1107.4123}]
.

\bibitem{arefeva1980}
I.Ya Aref'eva,
\emph{Quantum contour field equations},
contribution to \emph{17th Karpacz Winter School of Theoretical Physics: Fundamental Interactions},
\href{https://doi.org/10.1016/0370-2693(80)90529-8}
  {\emph{Phys. Lett. B} \textbf{91} (1980) 347-353}
.

\bibitem{gervaisnevau1979}
J.L. Gervais and A. Neveu,
\emph{The slope of the leading Regge trajectory in quantum chromodynamics},
\href{https://doi.org/10.1016/0550-3213(80)90397-1}
  {\emph{Nucl. Phys. B} \textbf{163} (1980) 189-216}
.

\bibitem{agliettieetal1992}
U. Aglietti, M. Crisafulli and M. Masetti,
\emph{Problems with the Euclidean formulation of heavy quark effective theories},
\href{https://doi.org/10.1016/0370-2693(92)90695-Z}
  {\emph{Phys. Lett. B} \textbf{294} (1992) 281-285}
.

%\cite{Aglietti:1993hf}
\bibitem{Aglietti:1993hf}
U.~Aglietti,
\emph{Consistency and lattice renormalization of the effective theory for heavy quarks},
\href{https://doi.org/10.1016/0550-3213(94)90231-3}
{Nucl. Phys. B \textbf{421}, 191-216 (1994)}
.
%doi:10.1016/0550-3213(94)90231-3
%[arXiv:hep-ph/9304274 [hep-ph]].
%32 citations counted in INSPIRE as of 29 Nov 2023

%\cite{Mandula:1990fit}
\bibitem{Mandula:1990fit}
J.~E.~Mandula and M.~C.~Ogilvie,
\emph{A Lattice implementation of the Isgur-Wise limit},
\href{https://doi.org/10.1103/PhysRevD.45.R2183}
{\emph{Phys. Rev. D} \textbf{45}, 2183-2187 (1992)}
.
%doi:10.1103/PhysRevD.45.R2183
%33 citations counted in INSPIRE as of 05 Dec 2023

\bibitem{mandulaogilvie1992}
J. Mandula and M. Ogilvie,
\emph{A lattice implementation of the Isgur-Wise limit},
\href{https://doi.org/10.1016/0920-5632(92)90303-A}
  {\emph{Nucl.Phys.B Proc.Suppl.} \textbf{26} (1992) 459-461}
.

\bibitem{Liu:2022nnk}
Y.Z. Liu,
\emph{Lecture notes on
transverse-momentum-dependent parton
distribution function and soft functions
in the large-momentum effective theory},
\href{https://doi.org/10.5506/APhysPolB.53.4-A2}
  {\emph{Acta Phys. Pol. B} \textbf{53}, 4-A2 (2022)},
\end{thebibliography}
\end{document}